\documentclass[a4paper]{spie}
\usepackage[latin1]{inputenc}
\usepackage{amsmath}
\usepackage{amsfonts}
\usepackage{amssymb}
\usepackage[]{graphicx}
\usepackage{color}
\usepackage{bbm}

\title{ADMIRE: a locally adaptive single-image, non-uniformity correction and denoising algorithm: application to uncooled IR camera}
\author{Y. Tendero \supit{a}, J. Gilles\supit{b}, \skiplinehalf
            \supit{a} Centre de Math�matiques et de Leurs Applications (CMLA)\\
	  �cole Normale Sup�rieure de Cachan - 61 av du Pdt Wilson 94235 Cachan Cedex France.\\
           \supit{b} Department of Mathematics - University of California Los Angeles\\
	  520 Portola Plaza, Los Angeles, CA 90095.          
           }
\authorinfo{Further author information: (Send correspondence to Y. Tendero)\\E-mail: tendero@cmla.ens-cachan.fr, Phone: +33 1 47 40 59 48\\
}
\begin{document}
\maketitle
\begin{abstract}
We propose a new way to correct for the non-uniformity (NU) and the noise in uncooled infrared-type images. This method works on static images, needs no registration, no camera motion and no model for the non uniformity. The proposed method uses an hybrid scheme including an automatic locally-adaptive contrast adjustment and a state-of-the-art image denoising method. It permits to correct for a fully non-linear NU and the noise efficiently using only one image.
We compared it with total variation on real raw and simulated NU infrared images.
The strength of this approach lies in its simplicity, low computational cost.  It needs no test-pattern or calibration and produces no ``ghost-artefact''.
\end{abstract}

\section{Introduction}
\label{sec:intro}  

Infrared (IR) imaging has proved to be a very efficient tool in a wide range of industry, medical, and military applications. Infrared cameras are used to measure temperatures, signatures, to perform detection, etc. %
However, the performance of the imaging system is strongly affected by the random spatial response of each pixel sensor. Under the same illumination the readout of each sensor is different. It leads to a structured noise resulting in a row or line pattern in the images (depending on the readout system). This ``noise'' is called fixed pattern noise and produces ``non uniformity'' in the observed images. Theses differences between sensor readout are due to mismatches in the fabrication process, among other issues \cite{book_electro_optical} and are stronger at longer wavelength such as in infrared imaging \cite{scribner1991infrared}. The readout of a pixel sensor is a non linear \cite{book_electro_optical} function of the incoming luminance. %
\begin{figure}[!h]  
\begin{center}
\includegraphics[width=5cm]{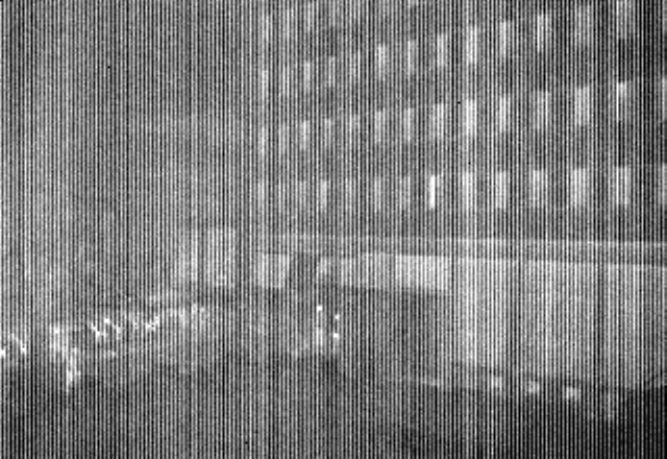}
\includegraphics[width=5cm]{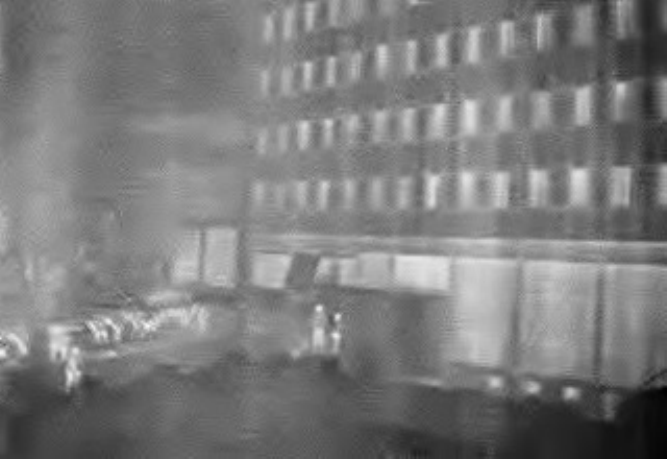}
\end{center}
\caption{\label{fig:figure1} On the left : an image (RAW) taken with an infrared camera. The non uniformity is so strong that it is hard to distinguish between the noise and the underlying landscape. On such an image performing an identification, matching pattern, etc. is almost hopeless. On the right the same image corrected with the proposed method.}
\end{figure}
The non uniformity is a serious practical limitation to both civilian and military applications - as it severely degrades image quality \cite{milton1985influence} (see Fig. \ref{fig:figure1}). For uncooled infrared cameras the problem is even worse because the detector response evolves quickly with time. Therefore the correction cannot be done once and for all by the manufacturer. It also means that we need to estimate, for each pixel, a function with little or no model at all and using little or one image to aim a good correction. Indeed the use of numerous images to achieve the correction leads to artefacts --those are called ``ghosts artefacts'' and are challenging to remove\cite{rossi2010temporal,Harris93minimizingthe,vera2010total,zuo2011scene,qian2011adaptive,zuo2011new}-- because of the sensor drift\cite{scribner1991infrared,scribner,riou2004nonuniformity}. In other words, for each pixel, the correction at times $t_1$ and $t_2\neq t_1$ are different. A correction is so much needed, that in many uncooled infrared cameras a flap closes every 30 seconds to perform a partial calibration \cite{jin2011infrared,houchin1991method}. This interrupts the image flow, which is calamitous for many applications. %
Thus, for uncooled infrared cameras a periodic update of the non uniformity correction is required. \\
A good non uniformity algorithmic correction is a key factor in ensuring the best image quality and the robustness of the downstream applications such as pattern recognition, image registration, etc. \\

In this paper we introduce locally adaptive version of our precedent work \cite{tendero:78340E}.
A review of existing techniques is proposed in  section \ref{sec:related}. %
A single image, fully automatic, non uniformity correction algorithm is detailed in section \ref{sec:previous_work} and generalized in section \ref{sec:admire}. %
It shows that motion compensation or accumulation algorithms are not necessary to achieve a good image quality. The proposed method can compensate for a fully non linear non uniformity, without any parametric model on the non uniformity side. It does not require motion, or motion compensation, does not need a test pattern or calibration and does not produce any ``ghost artifact''. %
A state of the art denoising algorithm is modified to suit our context.
The proposed method is illustrated in section \ref{sec:experiments} on simulated non linear non uniformity in section Figs. \ref{fig:simulated1}-\ref{fig:simulated2}, compared with a total variation based method (this method is described in section \ref{sec:total_variation_based_method}) in  section \ref{sec:comparative_experiments} and evaluated on real raw images from thermal infrared cameras in section \ref{sec:expe_real}.

\section{Image acquisition model}\label{sec:image_model}

An imaging sensor is a device that collects photons and converts them into charges. The majority of imaging sensors are Charge-Coupled Devices (CCD). The standard readout technique of CCDs works for each row (or line) independently and consists of transporting charges from the pixels to a counter (which produces the numerical value to be read). Each pixel has its own (and unknown) transfer function response. Furthermore, for each column the counter transfer function is different. The function resulting of the whole chain sensor-counter is not linear\cite{book_electro_optical}. In the sequel we will assume without loss of generality (w.l.o.g.) 
that the non uniformity comes only from the sensor part. %
It is the difference between (transfer) functions that produces the non uniformity and leads to a structured noise resulting in a row or line pattern in the images. 
The perturbation model is 
\begin{align} \label{eq:model}
o (i,j,t) = \phi_{(i,j,t)} \left( u (i,j,t) + \eta (i,j,t) \right), ~ \forall (i,j,t) \in \{1,...,N\}\times \{1,...,M\} \times \mathbb{R}_+,
\end{align}

where
\begin{itemize}
\item $(i,j,t) \in \{1,...,N\}\times \{1,...,M\} \times \mathbb{R}_+$ is the pixel at position $(i,j)$ and time $t \geq 0$;
\item $u(i,j,t)$ is the ideal (noiseless) landscape value;
\item $\eta (i,j,t)$ is a random photon noise;
\item $\phi_{(i,j,t)} :\{0,..., 255 \} \mapsto\{0,..., 255 \}$ is the contrast change (transfer function) of the pixel sensor at position $(i,j)$ and time $t \geq 0$;
\item $o (i,j,t)$ is the observed value at position $(i,j)$ and time $t \geq 0$.
\end{itemize}
Omitting the noise, $\phi_{(i,j,t)}( x )$ represents the readout of the pixel $(i,j)$ at time $t$ for some incident radiance $x$. At time $t$ the transfer function of the pixel $(i,j)$ and the transfer function of the counter are contained in $\phi_{(i,j,t)}$ (w.l.o.g). %
At each pixel sensor $(i,j)$ and time $t$ the photon (Poisson) noise $\eta (i,j,t)$ is sensed (and added to the ideal landscape pixel value $u(i,j,t)$), thus it also undergoes the contrast change function $\phi_{(i,j,t)}$. 
Consequently, (\ref{eq:model}) models thoroughly the whole acquisition process including the noise that is also modified by the non-uniformity of the sensor array. %
It permits to deals with a realistic \emph{non linear} sensor response \cite{book_electro_optical} contrarily to the classic linear (gain/offset) approximation used in the literature. %
The goal of a non-uniformity correction algorithm is to compensate for the \emph{local} contrast changes induced by the $\phi_{(i,j,t)}$ which means to apply some $\tilde \phi_{(i,j,t)}$ such that $\tilde \phi_{(i,j,t)}\left( \phi_{(i,j,t)}\right)=g(x)$ for $(i,j,t) \in \{1,...,N\}\times \{1,...,M\} \times \mathbb{R}_+$. Notice that, in general, it is useless to ask for $g=Id$ since users, screens (gamma correction), etc. usually tune the contrast of the images at ease.
\section{Related work}\label{sec:related}
To get the rid of the non-uniformity many techniques have been 
developed over the years. It is possible to classify them into two main kinds: 
\begin{itemize}
\item Calibration based techniques consist in an equalization of the response to an uniform black body source of radiations. They are not convenient for real time applications, since they force to interrupt the image flow. (This calibration is usually automatic, a shutter closing in front of the lens periodically). They usually assume that $\phi_{(i,j,t)}(x)=x+b_{(i,j)} ~\forall t \in [t_1,t_2[$ (so called one point NUC, one black body) or that $\phi_{(i,j)}(x)=a_{(i,j)}x+b_{(i,j)} ~ \forall t \in [t_1,t_2[	$ is linear (two points NUC using two black bodies). Thus it assumes piecewise constants $\phi_{(i,j,t)}$ on the time interval $[t_1,t_2[$. A new correction is performed every $t_2-t_1$ (because of the sensor drift). Of course the two points NUC \cite{friedenberg1998nonuniformity} performs better.
\item Scene based techniques, involving motion compensation or temporal accumulation. Such methods are complex and require certain observation conditions (motion). They usually assume linear $\phi_{(i,j)}$.
\end{itemize}
In the sequel, we will focus on scene based techniques as calibration based techniques require to interrupt the camera which is calamitous in practice. %
Numerous algorithms have been reported in the literature to remove the fixed pattern noise caused by the lack of a cross-pixels sensor equalization. %
Some algorithms estimate the sensor parameters while, equivalently, others attempt at recovering the ``true'' landscape value $u(i,j,t)$. These algorithms process a sequence of images $(o (i,j,t) )_{t \in {1,...,L}}$, not a single frame. Thus they are subject to the creation of ``ghost artefacts'', the reason is discussed bellow.  Most of them use a simplified (linear) model for the transfer function of the pixel sensor: 
\begin{align} \label{eq:model_approx}
o (i,j,t) = a_{(i,j,t)} \left( u (i,j,t) + \eta (i,j,t) \right) + b_{(i,j,t)}, ~ \forall (i,j,t) \in \{1,...,N\}\times \{1,...,M\} \times \mathbb{R}_+
\end{align}
There are methods like \cite{bb7630} suggesting to equalize the mean and standard deviation ($stddev$) \emph{through time} of each pixel sensor by a linear transform. Such algorithms rely on the data diversity found in most of the video sequences with some degree of motion. The key idea is

 \medskip [$\cal{H}$:] If all pixel sensors have seen the same landscape, they should have (at least) the same mean and same standard deviation, namely

\begin{align}
   mean \underset{t \in \left \{ 1,...L \right\}}{} \left( o (i,j,t) \right) &= C_m ~ \forall ~(i,j) \in \{1,...,N\}\times \{1,...,M\} \label{eq:syst1} \\
   stddev \underset{t \in \left \{1,...L \right\}}{} \left( o (i,j,t) \right) &= C_{std} ~ \forall ~(i,j) \in \{1,...,N\}\times \{1,...,M\}. \label{eq:syst2}
\end{align}
To summarize the authors suggest to adjust the sensor readout using a linear transform to enforce the equalities (\ref{eq:syst1}-\ref{eq:syst2}) above. But this is only possible if there is a long camera  sequence with enough motion where each sensor sweeps many different parts of the scene. Indeed a small window leads to little or no correction at all since it weakens [$\cal{H}$]. 
A subtle consequence of using long sequences is that the sensor (the contrast changes $\phi_{(i,j,t)}$) is assumed to be constant over the whole sequence. If the user cannot move sufficiently the camera, the convergence will be slow. This means that it is probable that between the beginning and the end of the sequence the $\phi_{(i,j,t)}$ will have changed because of the sensor drift.
It leads to an inaccurate estimation of $\phi_{(i,j,t)}$ : two pixels whom have not seen the same radiance should not satisfy (\ref{eq:syst1}) or (\ref{eq:syst2}). Moreover since the sensor is not linear, there will be some residuals. Theses residuals may are negligible except when the scene change suddenly: the approximation is wrong, because it is based on past observations. It will take time to update the estimation, depending on $L$. The residues of the correction as well as the previous landscape will remain superimposed in the subsequent frames. Those are the ``ghost artefacts''. The usual circumvent  to those ``ghost artefacts'' is to restart the learning process (forget some past data) if a new scene appears. Nevertheless, the detection of scene changes may be treacherous particularly if it occurs in a small portion of the image (a new vehicle, etc.) since it may be masked by the non uniformity (see Fig. \ref{fig:figure1}).\\
A variant like, for instance\cite{PezoaTCR04}, adjusts the minimum and the maximum of the readout values, assuming the time histograms observed in each sensor are equals over a long enough time sequence:

\begin{align}
   mean \underset{t \in \left \{ 1,...L \right\}}{} \left( o (i,j,t) \right) &= C_1~\forall~(i,j) \in \{1,...,N\}\times \{1,...,M\} \label{eq:syst3}\\
   stddev \underset{t \in \left \{1,...L \right\}}{} \left( o (i,j,t) \right) &= C_2 ~\forall ~(i,j) \in \{1,...,N\}\times \{1,...,M\}\label{eq:syst4}.
\end{align}

This last method is called Constant Range \cite{Torres02scene-basednon-uniformity}. As pointed out by several authors \cite{Harris93minimizingthe} the length  $L$ of the sequence is a crucial factor of success here. There is not way to tune $L$ {\it a priori} and two problems may arise:
\begin{itemize}
\item If $L$ is too small and the estimation is wrong because all sensors have not seen the same landscape ([$\cal{H}$] is wrong);
\item If $L$ is too large  and because of the  approximation bias and time drift of the sensor behavior, the previous images may appear as superimposed in the last ones. We retrieve the previously cited ``ghost artefact'' effect.
\end{itemize}
  There is a way to avoid the ``ghost artefacts'' \cite{Harris93minimizingthe}, which consists in a reset of the estimation when the scene changes too much. For example, in \cite{Harris93minimizingthe} the authors use a simple threshold to perform scene change detection (but the level of this threshold is not easy to tune in general). In \cite{rossi2010comparison} the author state that ``slow global motion and strong edges in the scene are the main causes" of the non uniformity. Indeed non consistent with [$\cal{H}$:] motions and/or bad length $L$ of the sequence used to enforce (\ref{eq:syst1}-\ref{eq:syst2} or \ref{eq:syst3}-\ref{eq:syst4}) lead to ``ghost artefacts'' because of the sensor drift. But, it is only more visible near edges.  Indeed the higher dynamic near edges weakens the linear approximation of the correction. The idea of treating edges separately  also appears, for example, in \cite{zhang2006edge}. %
To sum up, all these algorithms require a long exposition time with a varying scene and a serious (and sometimes involuntary)  camera motion.  \\

There are numerous implementations and studies \cite{scribner93,narendra1980reference,goyal2010nuc, qian2011adaptive,hart_thomas,Torres02scene-basednon-uniformity,hayat98,vera2003ghosting,vera2005fast} for these two major algorithms. A recursive filter \cite{bb7630} estimates the parameters of the linear function which approximates the $\phi_{(i,j,t)}$, or a Kalman filter is preferred \cite{Torres:03}. 
Other authors \cite{scribner91,nuc_torres_2003} propose a neural network based algorithm. Several other variants can be found in \cite{rossi2010comparison}.
 The registration based algorithms \cite{tzimopoulu98} consider often only translations\cite{cain97} (but homographies should be used instead, at least on a static scene). Creating a panorama has been proposed \cite{hardie00} to obtain a ground truth, and to use it as a calibration pattern. However, as pointed out \cite{DBLP:conf/icip/ZhaoZ06}, in presence of the structured fixed pattern noise occurring in  most infrared cameras, panoramas won't lead to a good result. Indeed a mean act as a low pass filter thus low frequencies of the non uniformity will remain in the produced images. \\
Recently in \cite{vera2010total} the authors minimize the total variation of the produced images. They also assume a linear model for the non uniformity. It generalizes \cite{moisan} but works on image sequences. Thus, it is also subject to ``ghost artefacts''.
In the following, the presented algorithm works on a single image. Furthermore, in the sequel we will omit the time dependence $t$ of the non uniformity ($\phi_{(i,j,t)}:=\phi_{(i,j)}$).
\section{Previous work: the midway infrared correction }\label{sec:previous_work}
The goal of this section is to give the background (section \ref{sec:midway}) and details (see sections \ref{sec:idea},\ref{sec:mire}) of our previous work \cite{tendero:78340E} (see \cite{ipol_mire} for an on line use and implementation in $C++$) as it is the starting point of the improvement presented here in section \ref{sec:admire}. 
\subsection{The midway histogram equalization method\label{sec:midway}}
\begin{figure}[!h]
\begin{center}
\includegraphics[width=12cm]{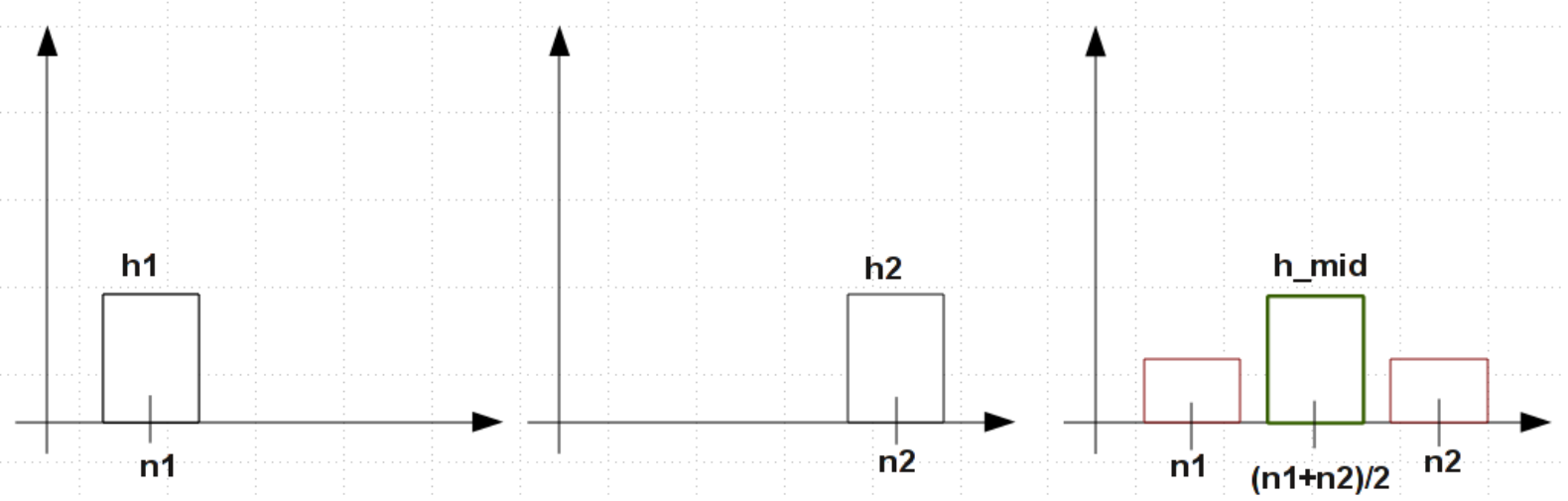}
\end{center}
\caption{\label{fig:mode}Two histograms $h_1$, $h_2$ (left side) and the corresponding midway histogram $h\_mid$  (on the right), compared to the direct histogram average, which would create two modes (centered at $n1$ and $n2$) and is therefore wrong. The uniform equalization would destroy the grey level dynamic and create artefacts. It is not a good candidate to get good quality images.}
\end{figure}

The midway algorithm was designed initially to correct for gain differences between cameras \cite{delon2004midway}. It permits to compare two images taken with different cameras more easily after their histograms have been equalized. This algorithm was later extended to flicker correction \cite{bb77071}. The midway equalization achieves much better and smoother results than giving flat (uniform) histograms to images. The idea of the midway equalization is to replace the (arbitrarily chosen) uniform histogram of classic uniform equalization by an histogram depending on the input images. It is optimal in the sense of the Wasserstein (transport) distance; the midway histogram being at equal distance of the histograms of input images. The midway equalization is defined and explained bellow.\\
 Consider two cumulative histograms $H_1$, $H_2$ of two images.  The midway cumulative histogram of the corrected image is simply  $$Hmid^{-1}:=\frac{H_1^{-1}+H_2^{-1}}{2},$$ and this average can be extended to an arbitrary number of images (and to non constant weights). A precise definition of the pseudo inverse is given in section \ref{sec:mire}. Once the midway histogram is computed, a monotone contrast change is applied to images to specify $Hmid$ as their common histograms. Thus, all images get the midway histogram, which is the best compromise between all histograms (see Fig \ref{fig:mode}).

\subsection{The idea\label{sec:idea}}
Since many infrared correction algorithms actually propose to equalize the temporal histograms of each pixel sensor, the midway is quite adapted to get a better result than a simple equalization. 
Equalization can be based on the fact that single columns (or lines, depending of the readout system) carry enough information by themselves for an equalization.
The images being continuous, the difference between two adjacent columns is statistically small, implying that two neighboring histograms should be nearly  equal. This hypothesis here is similar to the temporal one [$\cal{H}$] but is better suited to the decision to carry the equalization inside the image itself. It does not require any additional \emph{hypothesis} on the non uniformity (linearity, etc.).
In other words, the proposition is to transport  the histogram of each column (or line) to the midway of histograms of neighboring columns (resp. lines). In presence of strong fixed-pattern-noise (FPN) it will be useful to perform this {\it sliding midway method} over a little more than two columns, because the FPN is not independent in general.\\
\subsection{The midway infrared equalization algorithm (MIRE)\label{sec:mire}}
We give here the numerical details to implement the midway infrared equalization algorithm. It is fully automatic, compensate for non linear non uniformity thus, it is well suited to be used in an infrared image denoising chain (as a preprocessing for example).
Assume in the sequel that the equalization is performed among the columns of a discrete (8-bits w.l.o.g.) image $o (i,j) \in \{0,...,255\} ~ \forall (i,j) \in \{1,...,N\}\times \{1,...,M\}$. The ``midway infrared equalization (MIRE)'' algorithm proceeds as follows

For each column $j \in \{1,...,M\}$;
\begin{enumerate}
\item Compute the cumulative histogram $H_j$ of each column $ c_j$\\

$H_j(l)=\frac 1 N \sum_{k=0}^l \sum_{i=1}^N \mathbf{1}_{ \{o(i,j)=k\} }$;

\item For each column $c_j$ compute a local midway histogram $\tilde H_j^{-1}:=\sum_{k \in (-n,...,n)} g(k) H_{k+j}^{-1}$ using Gaussian weights $g(k)=g_s(k)=\frac{1}{s \sqrt{2 \pi}} e^{\frac{-k^2}{2s ^2}}$ with standard deviation $s$ and $n=round(4s)$ where,\\

$H_j^{-1}(l)=min \{ z \in \{0,...,255 \} /H_j(z) \geq l \}$ ;

\item Specify the histogram of the column $c_j$ onto this midway histogram $\tilde H_j$\\

$d(i,j)= \tilde H_j^{-1} \left( H_j(o(i,j)) \right) ~ \forall i\in \{1,...,N \}$.
\end{enumerate}
Since we work on images separately the method is not affected by motions or scene changes. This completely avoids ``ghost artefacts'' \cite{Harris93minimizingthe,rossi2010temporal,hardie2009scene} and any problem caused by the calibration parameters drifting over time. Notice that contrarily to the classic literature algorithms it does not assume any additional properties (linearity, etc.) on the non uniformity. The histogram specification step is also non linear in general. Thus, we can compensate for fully non linear non uniformity. 
An algorithm selecting $s$ is given in the next paragraph. 

\subsubsection{Automatically fitting the perfect $s$ parameter\\}\label{sec:s_param}
The non-uniformity leads to an increased total-variation norm. Hence, following the idea of \cite{moisan}, the smoothest image is also the one with little or no non-uniformity at all. So the simplest way to find the good ($s^*$) parameter automatically is :
$$s^*=argmin_s ||I_s||_{TV-line}$$ where $I_s$ is the image processed by the MIRE algorithm described above with the parameter $s$. The discrete (line) total variation is defined by $||I||_{TV-line}=\sum_{i,j}|(\nabla I)_{i,j}| , (\nabla I)_{i,j}= \left( I_{i+1,j} -I_{i,j} \right)$.
The optimization can be done by scanning a broad range of $s$. Choose a $s\_step$ and a $s\_max$ ($s\_step=0.5$ and $s\_max=8$ by default in \cite{ipol_mire} and never lead to unsatisfactory results). Start with $s=0 $, repeat : 1) process the image 2) increase $s$ of $s\_step$, then stop when $s>s\_max$. \\

\begin{figure}[!h]  
\begin{center}
\includegraphics[width=5cm]{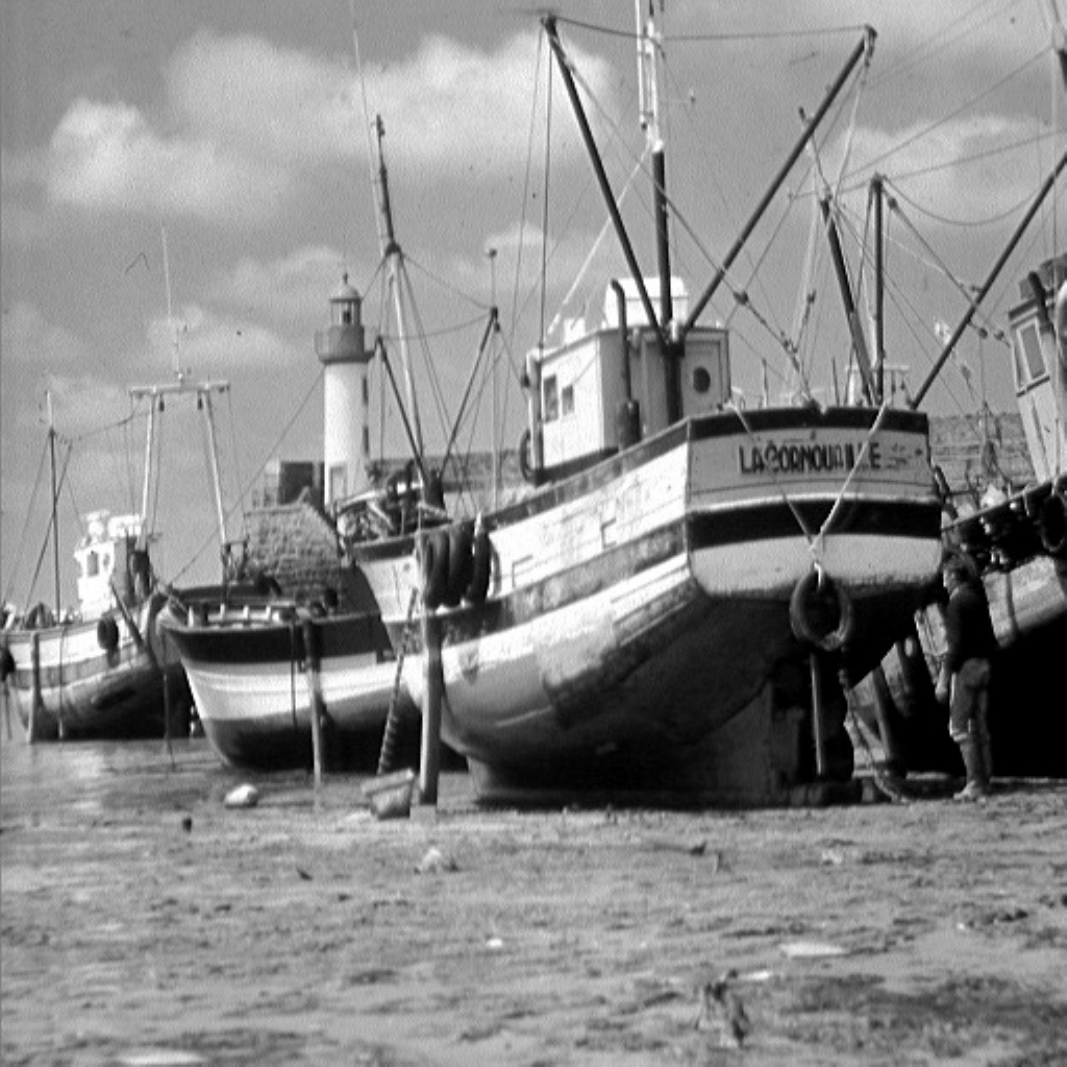}
\includegraphics[width=5cm]{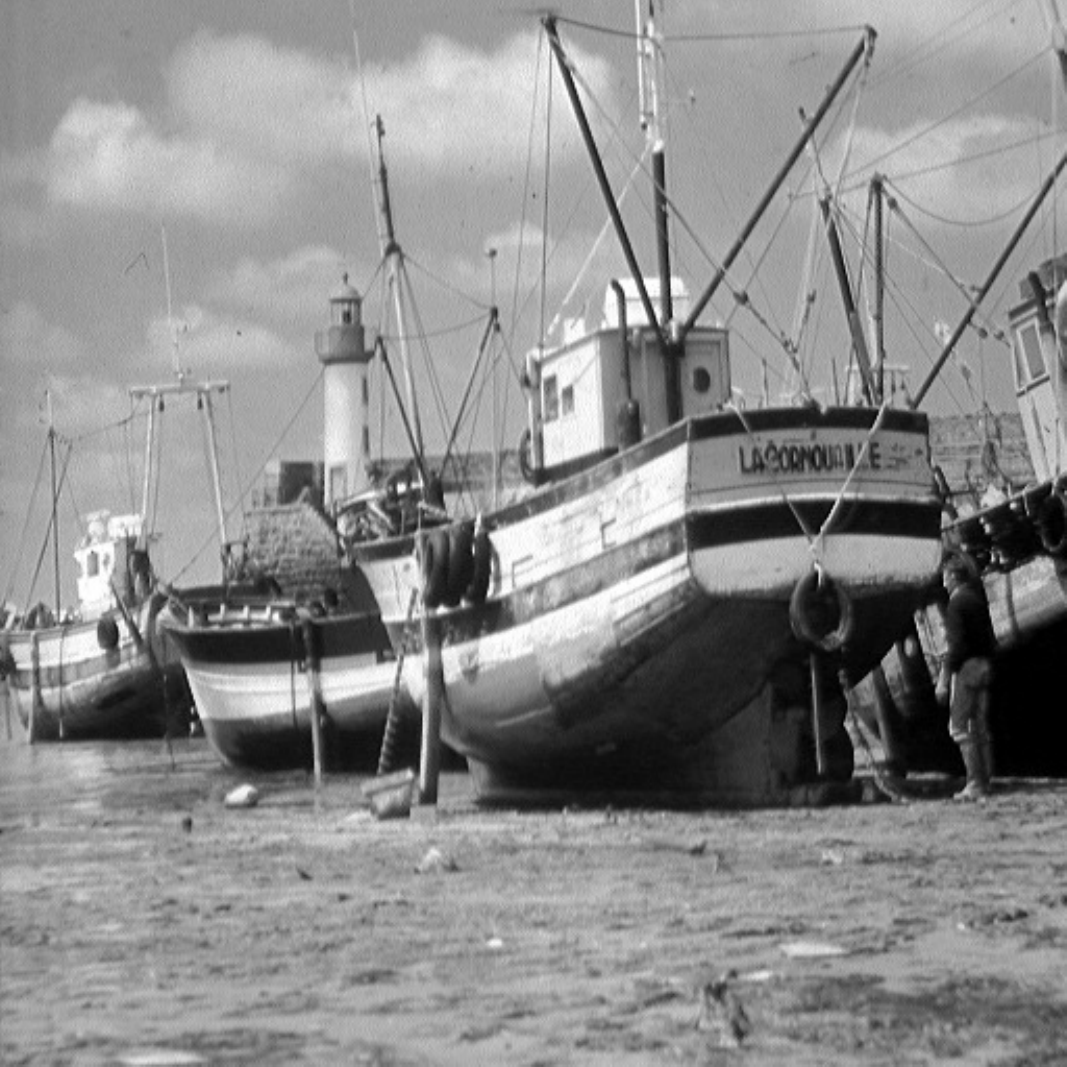}
\includegraphics[width=5cm]{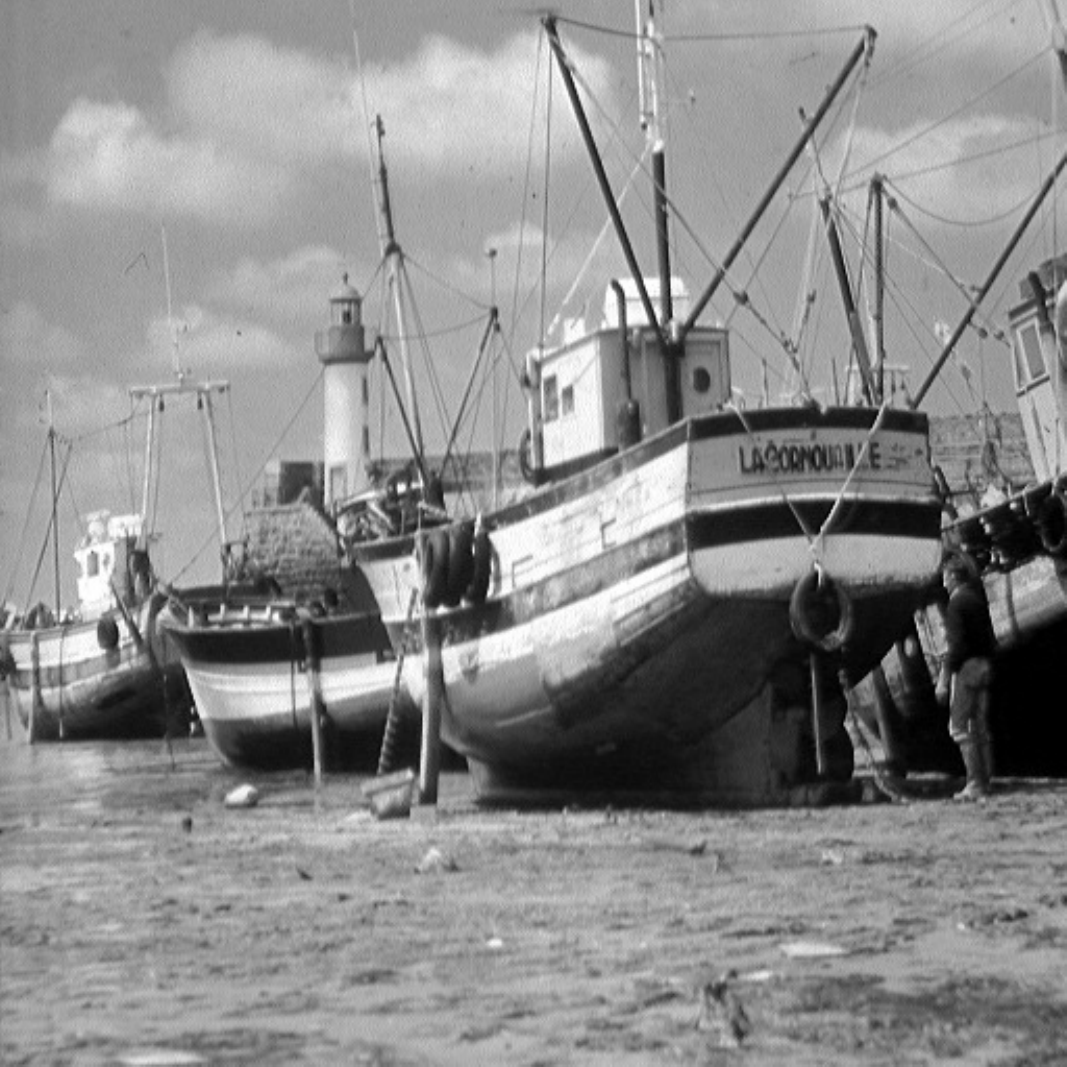}
\end{center}
\caption{\label{fig:safety} On the left : an uncorrupted test image (boat). On the middle : the result of the MIRE algorithm ($s^*=0$); the produced image is the same. This experiment was done using \cite{ipol_mire}. On the right : the result of the locally adaptive variant of MIRE described in section  \ref{sec:adaptive} ($s^*=0$ everywhere in the image). As predicted the algorithms does not make the image worse or create artefacts (safety check). Results on real raw images corrupted with non uniformity are detailed in section \ref{sec:experiments}.}
\end{figure}

The following ensures a certain safety of the proposed method, a fact that is confirmed by the experience of Fig \ref{fig:safety}.\\
\noindent{\bf Theorem 1.}
If  $H_i$ $i \in \{-n,...,n\}$ are $2n+1$ cumulative histograms of the same landscape seen by $2n+1$ different columns of the sensor, and $ H_{mid}^{-1}=\sum_{j=-n}^{n} \frac{H_{j}^{-1}} {2n+1} $ then :
$$ ||H_{mid}-H_{true}||_2 \leq max\underset{i \in \{-n,...,n\} }(||H_i-H_{true}||_2)$$
Moreover if the $H_i ~ \forall i \in \{-n,...,n\}$ from the $2n+1$ columns of the sensor are \textit{i.i.d. and centered} on $H_{true}$ then $$||H_{mid}-H_{true}||_2\underset{n \to \infty }{\rightarrow}0$$\\

\section{The Adaptive and Denoising midway equalization algorithm (ADMIRE)}\label{sec:admire}
\begin{figure}[!h]  
\begin{center}
\includegraphics[width=8cm]{./processed/test9_0_0003}
\includegraphics[width=8cm]{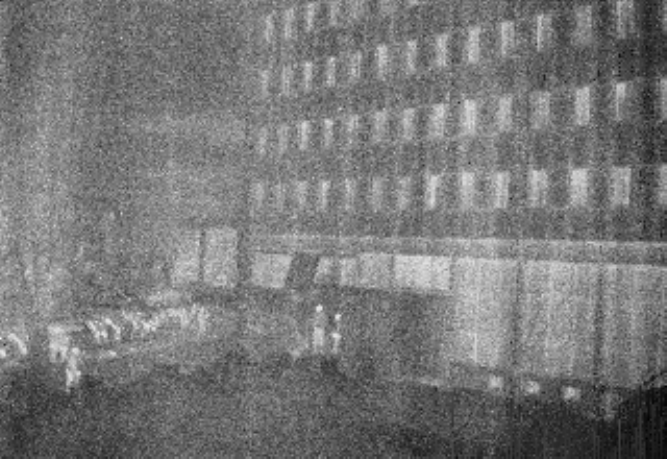}
\end{center}
\caption{\label{fig:noisy} On the left : a real raw image produced by an infrared camera. On the right : the result of the locally adaptive MIRE algorithm (see section \ref{sec:adaptive}). It is strongly corrupted by noise.}
\end{figure}

In this section we describe the novelties proposed to our previous work (see section \ref{sec:previous_work} and \cite{tendero:78340E,ipol_mire}). It consists in a modification of the MIRE (see section \ref{sec:mire}) to make it locally adaptive to the image. The need for a locally adaptive scheme is illustrated bellow (see Fig. \ref{fig:adaptive}). This modification is detailed in section \ref{sec:adaptive}. The result of this locally adaptive scheme is nevertheless corrupted by the noise (which may be strong as in Fig. \ref{fig:noisy}). Thus, we shall also embed a denoising scheme in order to increase the signal to noise ratio. 
\subsection{The adaptive midway equalization algorithm}\label{sec:adaptive}
\begin{figure}[!h]  
\begin{center}
\includegraphics[width=5.5cm]{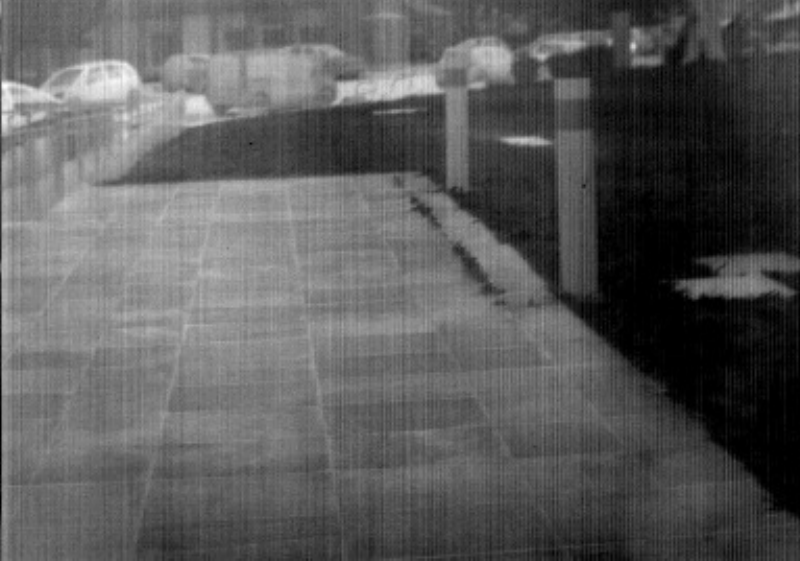}
\includegraphics[width=5.5cm]{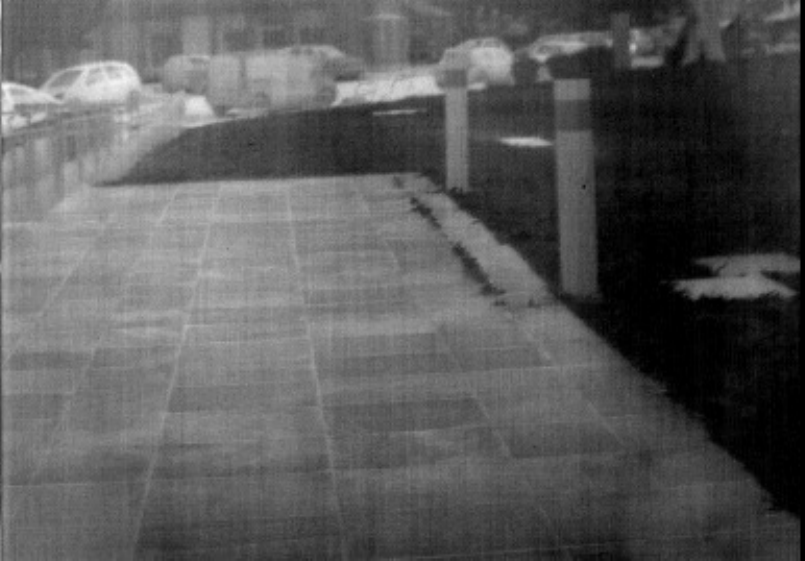}
\includegraphics[width=5.5cm]{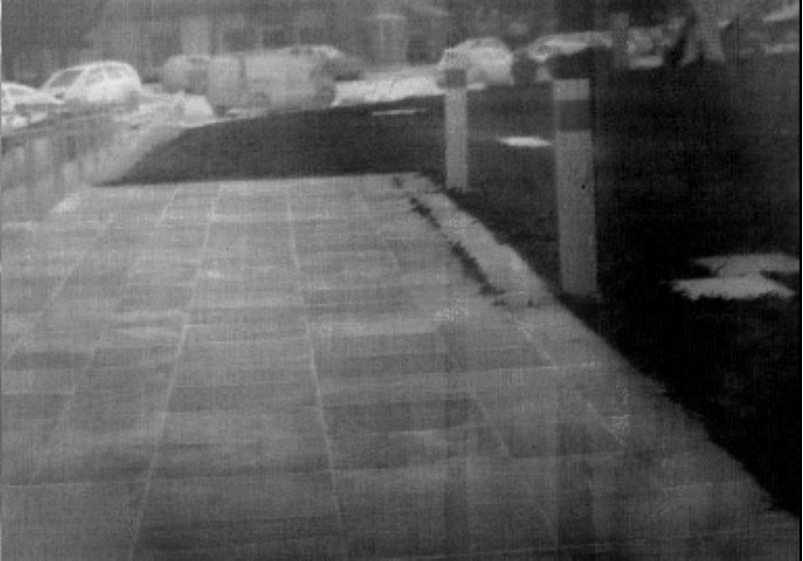}
\end{center}
\caption{\label{fig:adaptive} On the left : a real raw image produced by an infrared camera. Middle : the result of the algorithm with the parameter $s=1$. On the right with $s=7.5$. For example, focusing on a zone in the middle of the image the image with $s=7.5$ is nicer but, the zone below poles is bad. On the other hand in the image processed with $s=1$ the zone in the middle is still corrupted by the non uniformity. Thus, a fixed $s$ parameter for the whole image will not lead to the best quality possible everywhere.}
\end{figure}

There is no real justification (except simplicity) to keep a constant parameter $s$ over the whole image. Indeed the image can present different contents and structures in different areas. Thus focusing on different part of the image the best $s$ parameter may be different. A fact that is illustrated in Fig. \ref{fig:adaptive}. Thus, we propose to adapt the $s$ parameter of the MIRE algorithm locally. The proposed algorithm is :
\begin{enumerate}
\item For each $s$ in $s\_min:s\_max$ (see section \ref{sec:s_param}) process the image by MIRE;
\item Decompose all images in patches ($8\times8$ patches always used here);
\item For each patch keep the one with smallest $TV-line$ (the best one as in section \ref{sec:s_param});
\item Average and aggregate all patches to get the non uniformity corrected image.
\end{enumerate}

\subsection{The denoising step : anisotropic $DCT$ threshold on overlapping patches}\label{sec:denoising}
Since the processed images seem to be corrupted by a strong noise we propose to adapt \cite{coifman1995translation} to perform a good denoising. 
Thus we propose to modify the $DCT$ threshold denoising algorithm of \cite{coifman1995translation} which performs a $DCT$ threshold on sliding and overlapping patches. It is well suited with the previous step of section \ref{sec:admire} which also use patches. Moreover \cite{coifman1995translation} is unaffected\footnote{The non local means denoising \cite{buades2005non} has issues on some images (depending on the amount of residues of the non uniformity). Indeed residues of non uniformity interfere with patch distances (even after a column by column normalization of patches by their variance.)} %
by residues of non uniformity. %
We kept $8\times 8$ patches as in \cite{coifman1995translation} for the examples given below in section \ref{sec:experiments}. The  modification consists of the use of two different thresholds in the direction of the non uniformity and the orthogonal direction. Indeed the coefficients in the direction of the lines should be bigger (for an image corrupted by columns non uniformity noise as in Fig. \ref{fig:figure1}) to denoise more. Indeed this direction is more corrupted by residues of non uniformity. It leads to an anisotropic filtering of the patches. To sum up, we suggest to perform an anisotropic $DCT$ threshold on overlapping patches as a final denoising step. Provided two thresholds $T_i$ and $T_j$ (denoising strongness) in the $i$ (lines being indexed by $i$) and $j$ direction of the image the algorithm is :

\begin{enumerate}
\item Decompose the image sliding patches;
\item For each patch :
\begin{enumerate}
\item Compute 2D-DCTII transform of the patch;
\item Threshold the DCTII coefficients, with a threshold equal to $T_j$ in the $j$ direction and $T_i$ (lines being indexed by $i$) everywhere else;
\item Calculate inverse 2D-DCT transform of the patch;
\item (normalize by a factor of $\frac{1}{4 patch\_size*patch\_size}$);
\end{enumerate}
\item  Average and aggregate all patches to get the denoised image.
\end{enumerate}
\subsection{Implementation}
The implementation is easy and was done with \textit{C++}. To avoid border effects we used a reflection of the image across borders. A $C++$ source code of the MIRE algorithm is available in \cite{ipol_mire}, it allows on line experiments. The demo performs the MIRE algorithm, permits to see and download the result and shows the $s^*$ parameter computed by the algorithm. For the denoising step we refer to \cite{ipol.2011.ys-dct}, where a $C++$ source code is available (as well as an on line demo). %
The overall chain of ADMIRE is :
\begin{enumerate}
\item Perform the adaptive midway equalization algorithm described in section \ref{sec:adaptive};
\item Perform the anisotropic $DCT$ threshold on overlapping patches described in section \ref{sec:denoising}.
\end{enumerate}
Of course a temporal extension (to videos) of the proposed method avoiding flicker (and ``ghost artefacts'') is possible, using a temporal midway \cite{bb77071}.

\subsection{\label{remark:rmse_ci}Quality analysis; contrast invariant $RMSE$}
In many cases reconstruction errors inherent to a method can be quantified using
the Root-Mean-Squared-Error $ RMSE \left( u,u_{est} \right)
:=\sqrt{\frac{\int_D|u(x)-u_{est}(x)|^2dx}{measure(D)}} .$ However, when a small
contrast change occurs between the original and the processed image, the $RMSE$ can
become substantial, while the images remain perceptually indistinguishable. For
example take an image $u(m,n)$ defined over a sub-domain $ D \subset\mathbb{Z}^2 $,
and another one $u_{est}=u+10$. Then $RMSE(u,u_{est})=10$  is large but does not
reflect the quality of the reconstruction $ u_{est}$. Comparatively, a convolution
with, for example, a Gaussian can give a smaller $RMSE$ while making considerable
damage.
This bias is avoided by normalizing the images before computing the $RMSE$. The principle of the normalization is that two images related to each other by a contrast change are perceptually equivalent. Their distance should reflect this fact and be zero. The midway equalization \cite{delon2004midway} is best suited for that purpose, because it equalizes the image histogram to a ``midway'' histogram depending on both images. %
By the midway operation both images undergo a minimal distortion and adopt exactly
the same histogram. Thus we shall define the contrast invariant $RMSE$
($RMSE^{CI}$) by
\begin{align*}
RMSE^{CI}=RMSE(u_{{est}_{mid(u,u_{est})}},u_{mid(u,u_{est})})
\end{align*}
where $mid(u,u_{est})(=mid(u_{est},u))$ is the midway histogram between $ u $ and
$ u_{est} $.  $ u_{mid(u,u_{est})}$ is the image  $u$ specified on the $
mid(u,u_{est}) $ histogram (having an histogram equal to  $  mid(u,u_{est})$) and $
u_{est}{_{mid(u,u_{est})}} $ is  $u_{est} $ specified on $ mid(u_{est},u) $.

\section{Experiments}
\label{sec:experiments}
\begin{figure}[!h]
\begin{center}
\includegraphics[width=8cm]{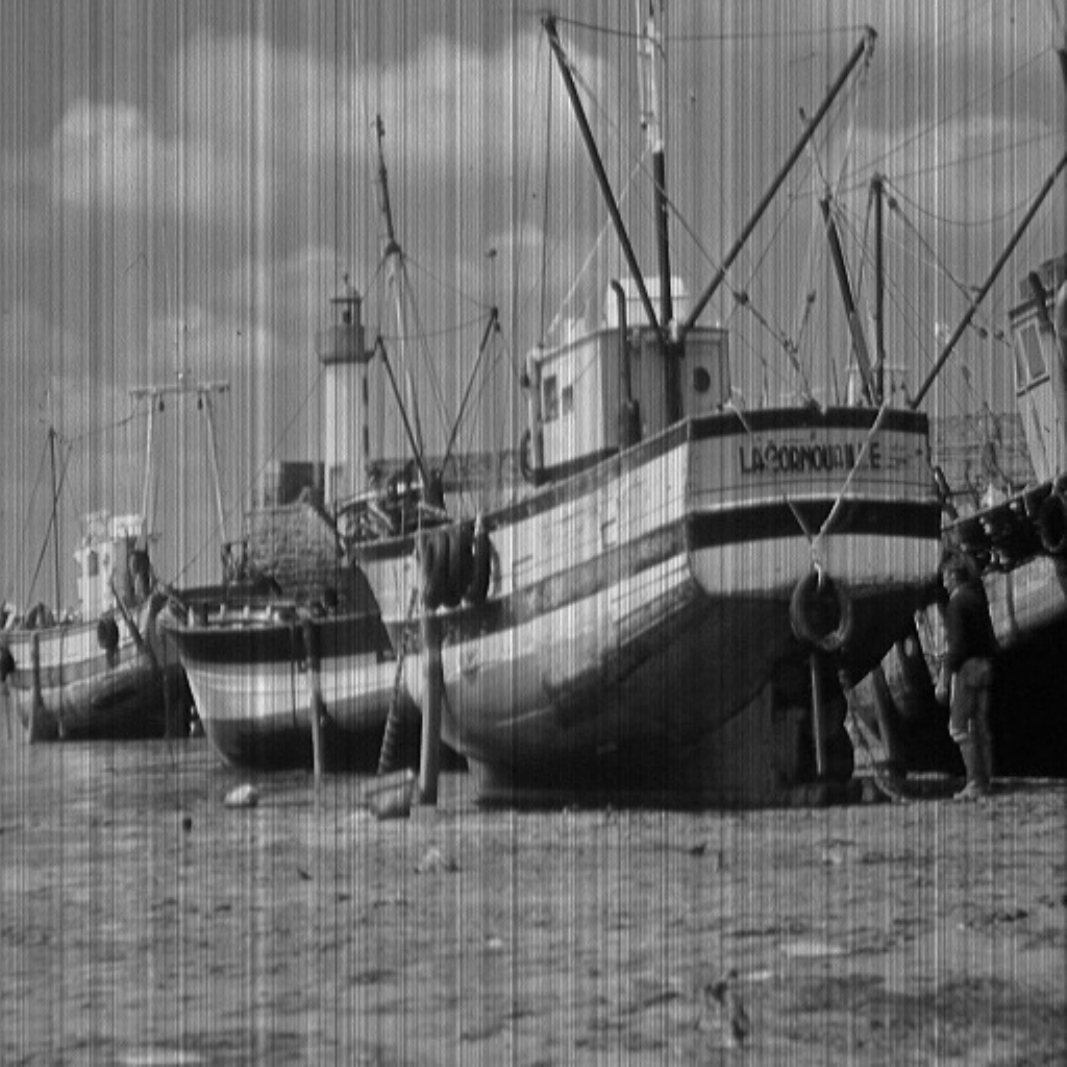}
\includegraphics[width=8cm]{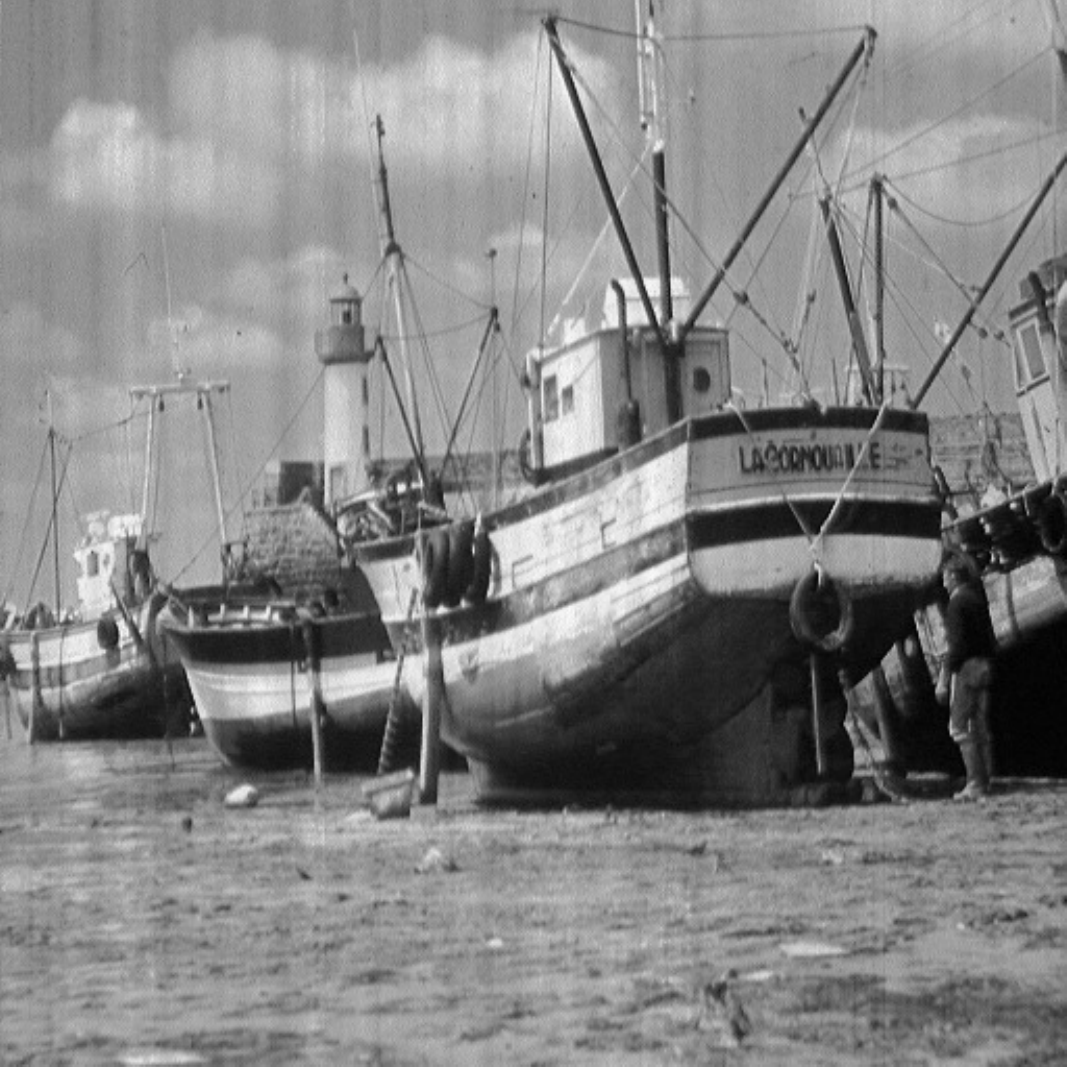}
\caption{\label{fig:simulated1}On the left : an image with a strong simulated non linear non uniformity. On the right : the image processed by the proposed algorithm. $RMSE=9.6629$, $RMSE^{CI}=5.7314$.}
\end{center}
\end{figure}

\begin{figure}[!h]
\begin{center}
\includegraphics[width=8cm]{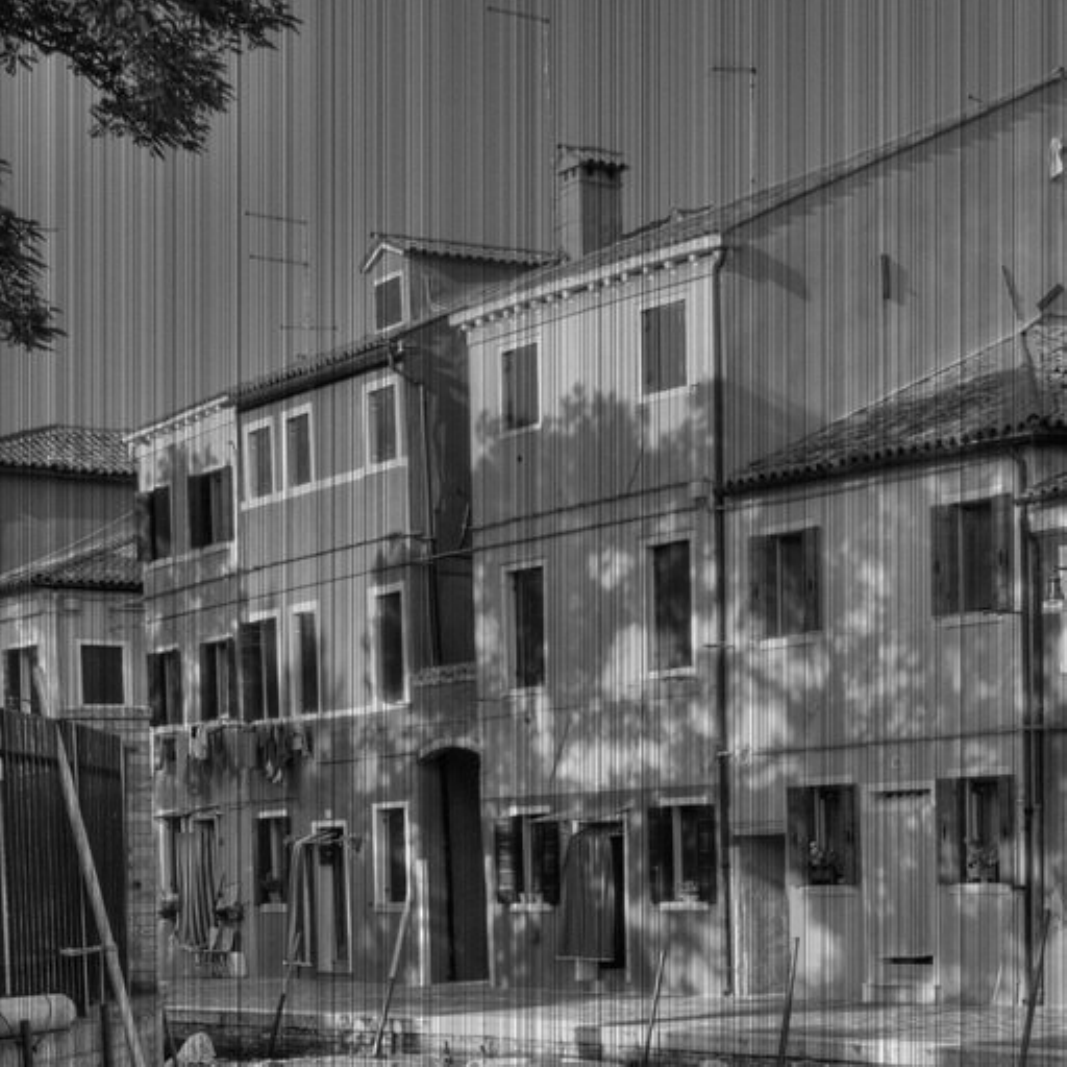}
\includegraphics[width=8cm]{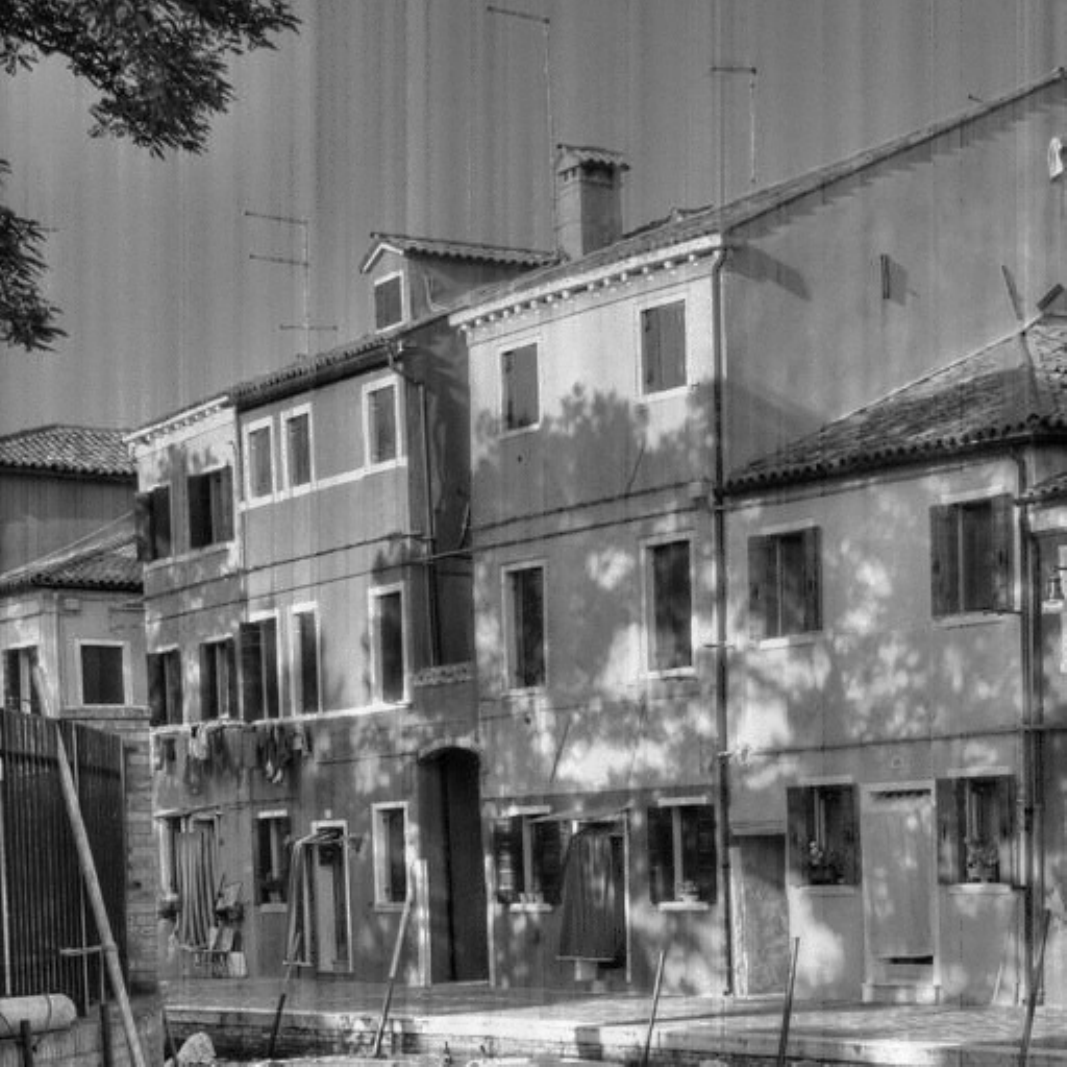}
\caption{\label{fig:simulated2}On the left : an image with a strong simulated non linear non uniformity. On the right : the image processed by the proposed algorithm. $RMSE=8.7100$, $RMSE^{CI}=7.3411$.}
\end{center}
\end{figure}

Simulations of Fig. \ref{fig:simulated1}-\ref{fig:simulated2} are made using a nonlinear randomly generated model of NU. Results are quantified in term of $RMSE$ and $RMSE^{CI}$ (the contrast invariant $RMSE$ reflects more the intrinsic quality of the image). They confirm the guess of visual improvement in quality. The comparisons with the total variation based method is given in Figs. \ref{fig:comp_tv1}-\ref{fig:comp_tv2}. The proposed algorithm outperforms it. The algorithm was run on real raw images, Figs. \ref{fig:expe1}-\ref{fig:expe6}, thus for theses experiments it is not possible to compute a $RMSE$ (as the groundtruth is unknown). The proposed algorithm outperforms classic literature methods in real situation on real (non simulated) non uniformity while using only one image.
\subsection{Total variation based method}\label{sec:total_variation_based_method}
Let $o (i,j) ~ \forall (i,j) \in \{1,...,N\}\times \{1,...,M\}$ be the observed image. The $TV-line$ based method \cite{moisan} looks for a constant $k(j)$ to add to each column. So $$|| o(i,j) +k(j)||_{TV-line}$$ is as small as possible. This boils down to the minimization of  $\sum_i|o(i,j+1)+\delta(j)-o(i,j)|$ for each column $j$ (notice that this sum involves only the column $j$ and its neighboring column $j+1$). Then $k(j+1)=k(j)+\delta(j)$, where $k(0)=c$ chosen so that the resulting image has the same mean as the observed image $o$. In practice this can be done by :
\begin{enumerate}
\item Keep the first column intact ($j=0$);
\item For each $j \in \{1,...,M\}$;
\begin{enumerate}
\item Minimize $\sum_{i \in \in \{1,...,N\}} |o(i,j+1)+\delta(j)-o(i,j)|$, by trying all possible $\delta(j)$ constants (using the quantification of the image);
\end{enumerate}
\item Add a constant to the whole image so the output has the same mean as $o$.
\end{enumerate}
%
%

\subsection{Comparative experiments}\label{sec:comparative_experiments}
\begin{figure}[!h]
\begin{center}
\includegraphics[width=5.5cm]{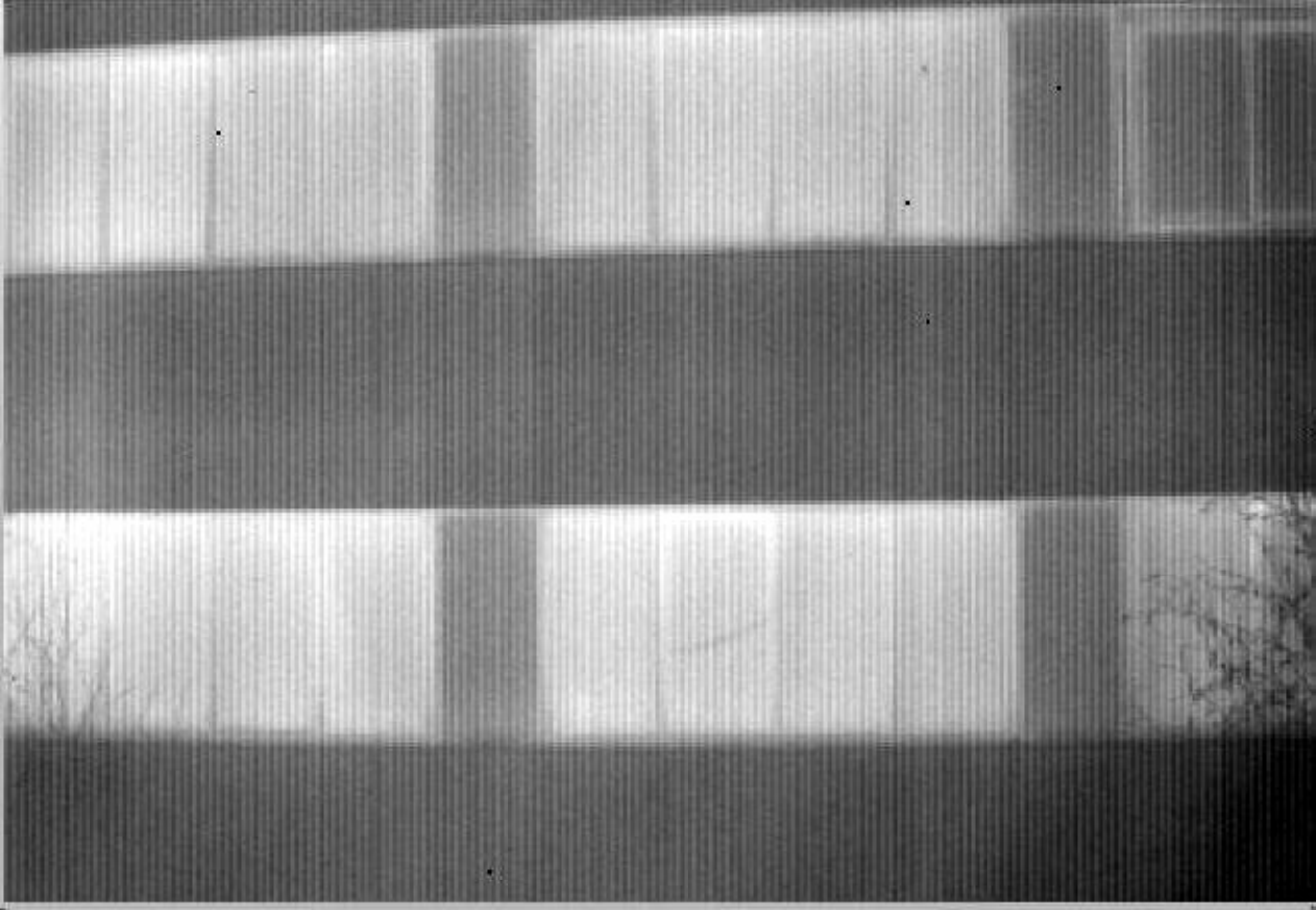}
\includegraphics[width=5.5cm]{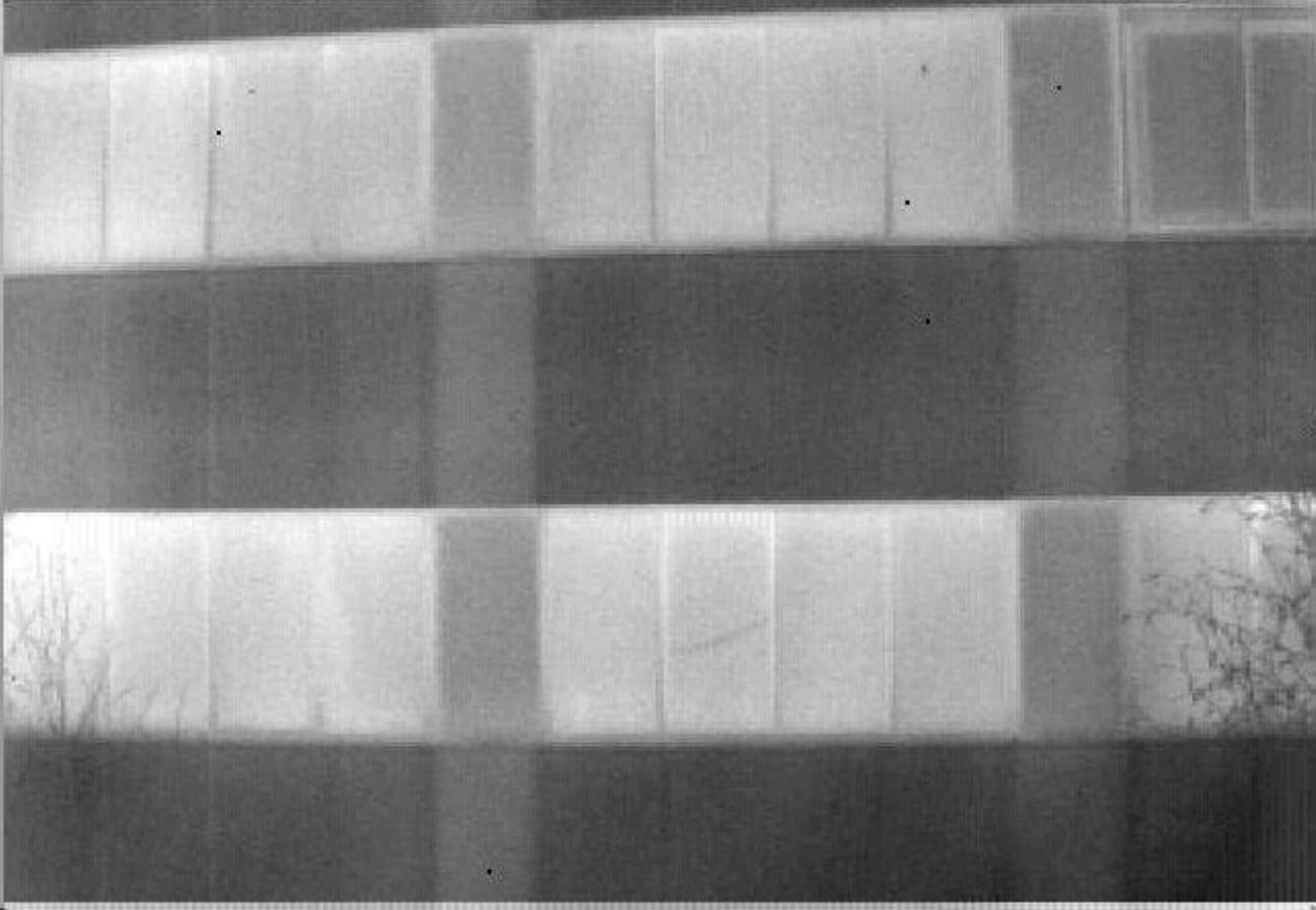}
\includegraphics[width=5.5cm]{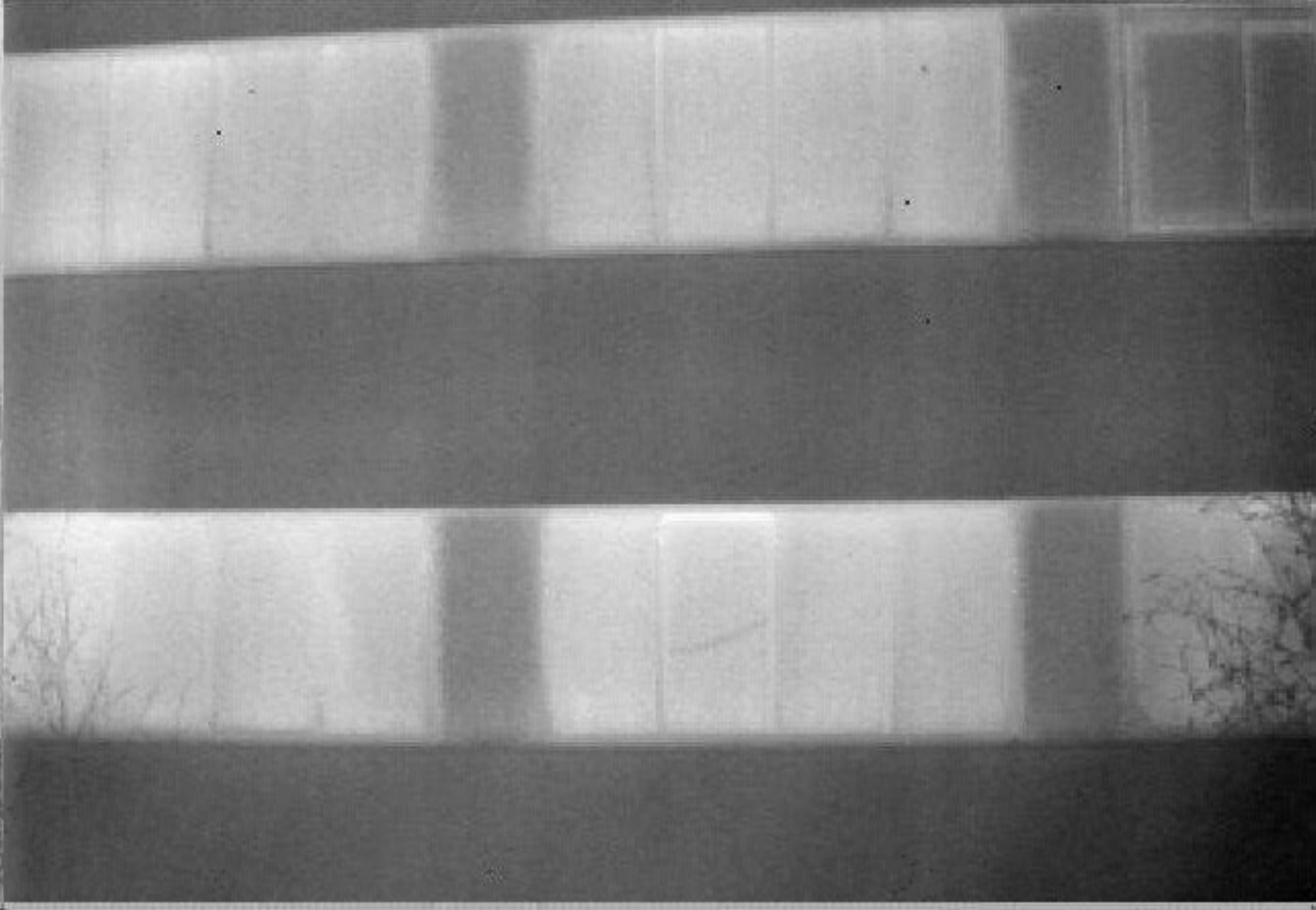}
\caption{\label{fig:comp_tv1}On the left : a real raw image of a building taken by an infrared camera. Middle : the total variation based method. Notice the artefacts created on the concrete stripes. The concrete stripes are at constant temperature thus, it should be at a constant grey level. On the right the proposed method (we deactivate the $DCT$ denoising to provide a fair comparison).}
\end{center}
\end{figure}

\begin{figure}[!h]
\begin{center}
\includegraphics[width=5.5cm]{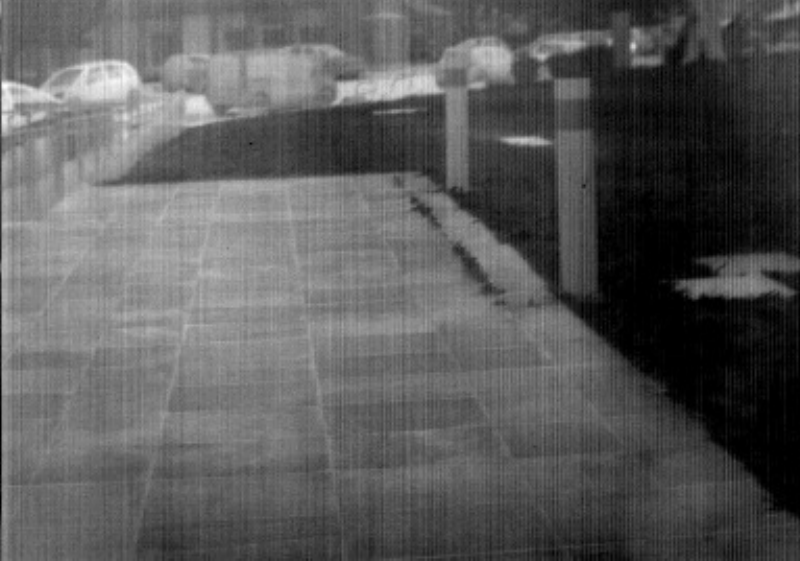}
\includegraphics[width=5.5cm]{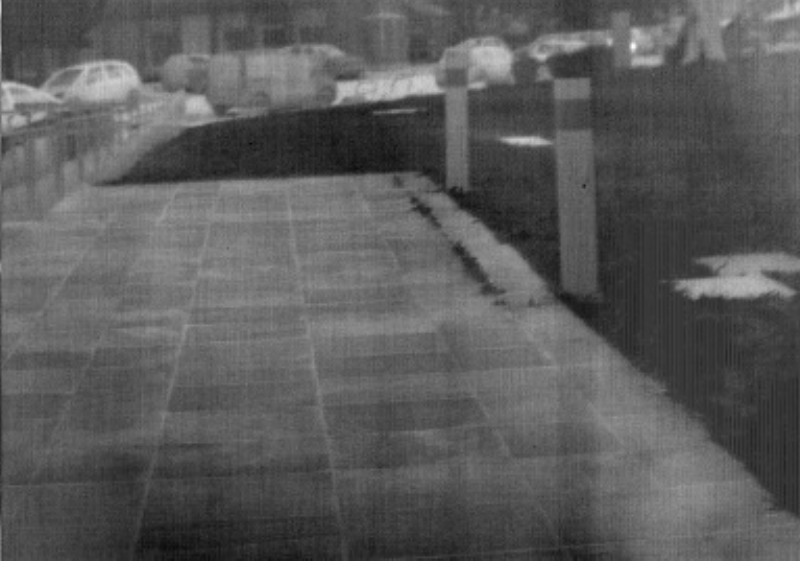}
\includegraphics[width=5.5cm]{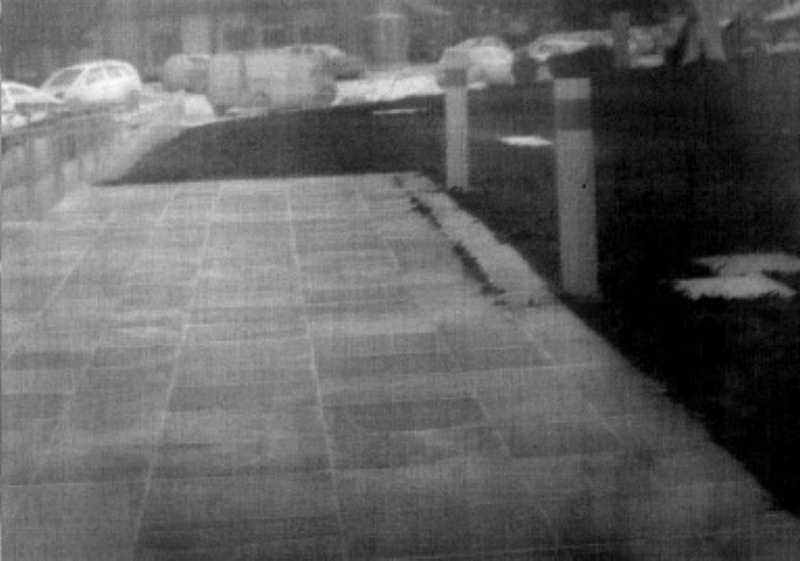}
\caption{\label{fig:comp_tv2}On the left : a real raw image of an outdoor scene taken by an infrared camera. Middle : the total variation based method. Notice the artefacts created bellow the poles. On the right the proposed method (we deactivate the $DCT$ denoising to provide a fair comparison).}
\end{center}
\end{figure}

The comparative experiments with the total variation were processed using \textit{Megawave \footnote{ \textit{Megawave} is available at megawave.cmla.ens-cachan.fr/  }} (resthline  module \cite{moisan}). 
Experiments on real raw infrared images are shown in Figs. \ref{fig:comp_tv1}-\ref{fig:comp_tv2}. The denoising step of ADMIRE has been deactivated for these experiments to allow a fair comparison with the total variation (TV) based method (which does not denoise). ADMIRE always shows a significant improvement on the TV based method and the final visual quality is very satisfactory. 

\subsection{Experiments on real raw images}\label{sec:expe_real}
\begin{figure}[!h]
\begin{center}
\includegraphics[width=8cm]{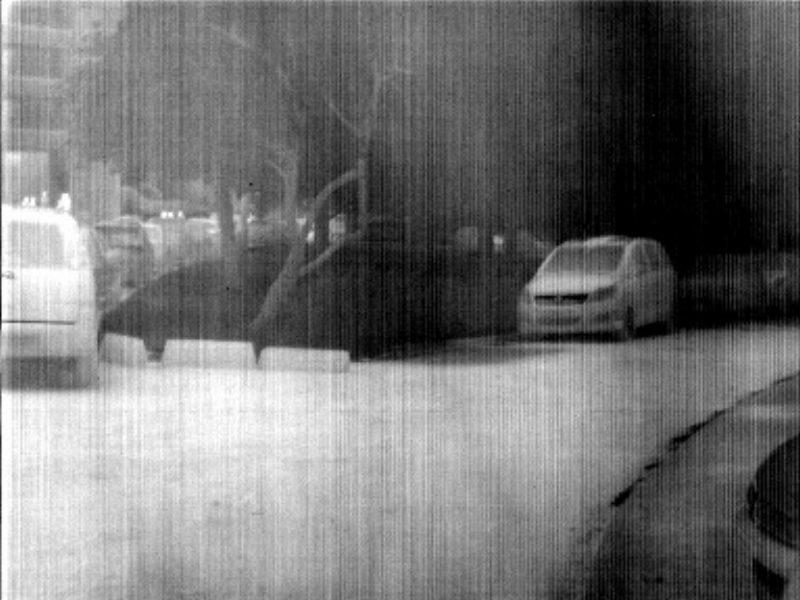}
\includegraphics[width=8cm]{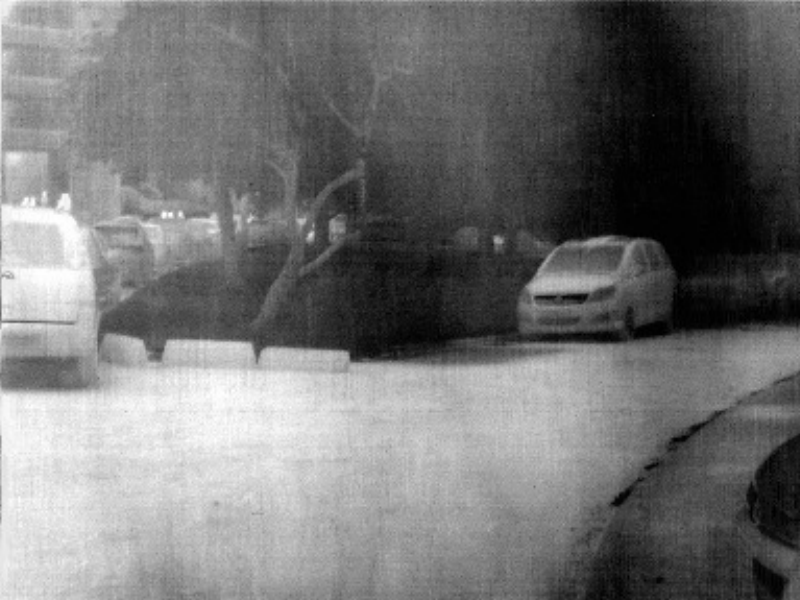}
\caption{\label{fig:expe1} On the left : a real raw image taken by an infrared camera. On the right the result of ADMIRE.}
\end{center}
\end{figure}

\begin{figure}[!h]
\begin{center}
\includegraphics[width=8cm]{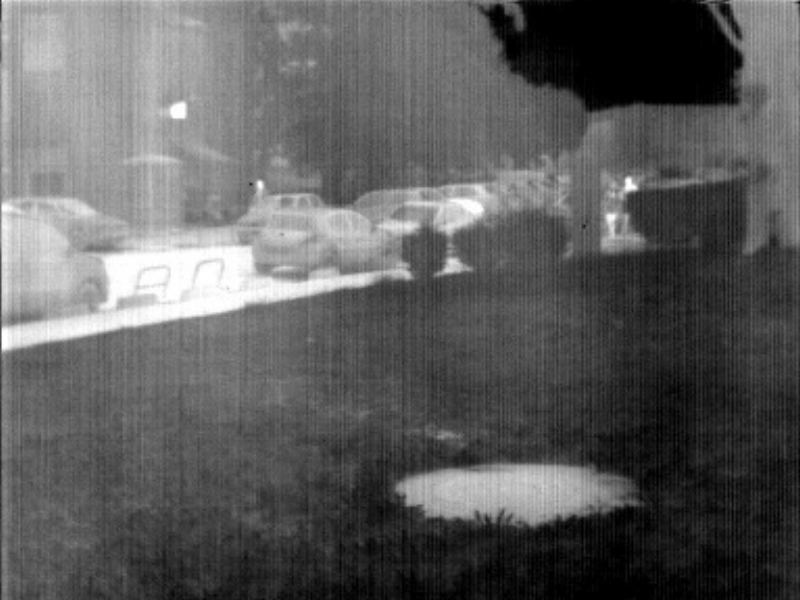}
\includegraphics[width=8cm]{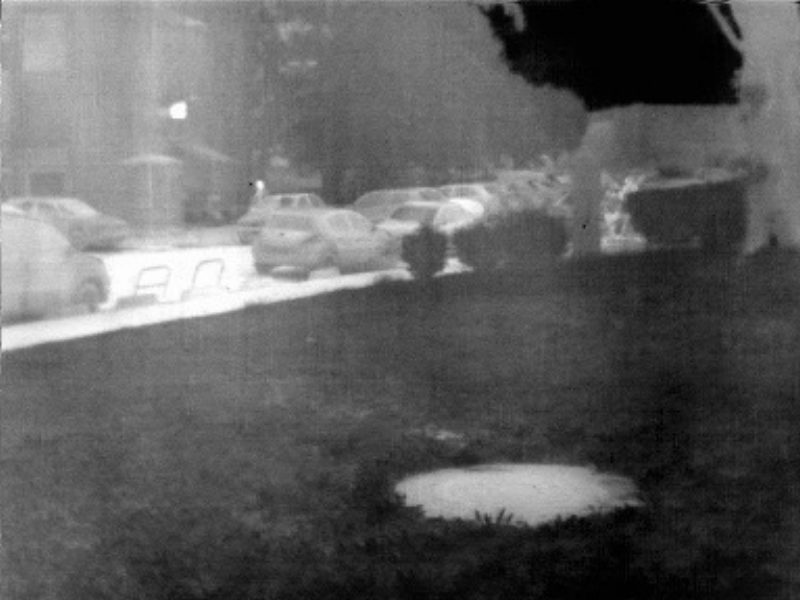}
\caption{\label{fig:expe2} On the left : a real raw image taken by an infrared camera. On the right the result of ADMIRE.}
\end{center}
\end{figure}

\begin{figure}[!h]
\begin{center}
\includegraphics[width=8cm]{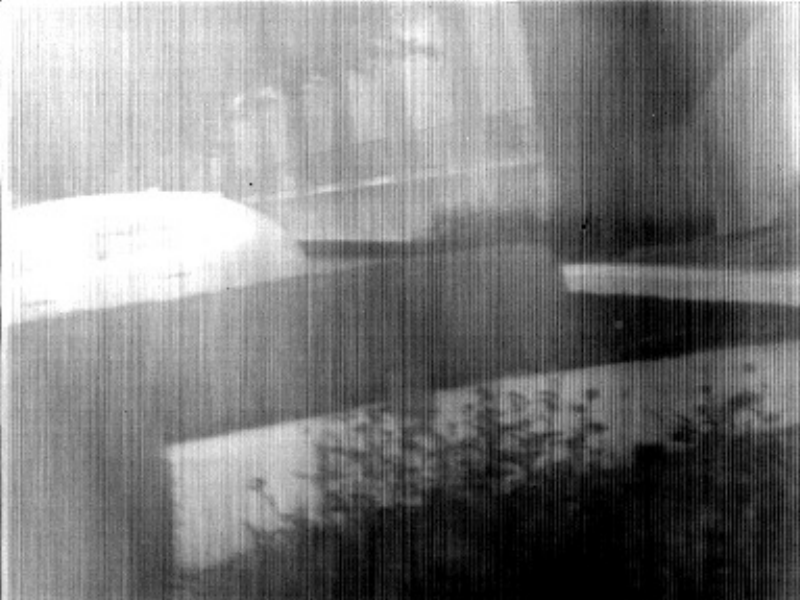}
\includegraphics[width=8cm]{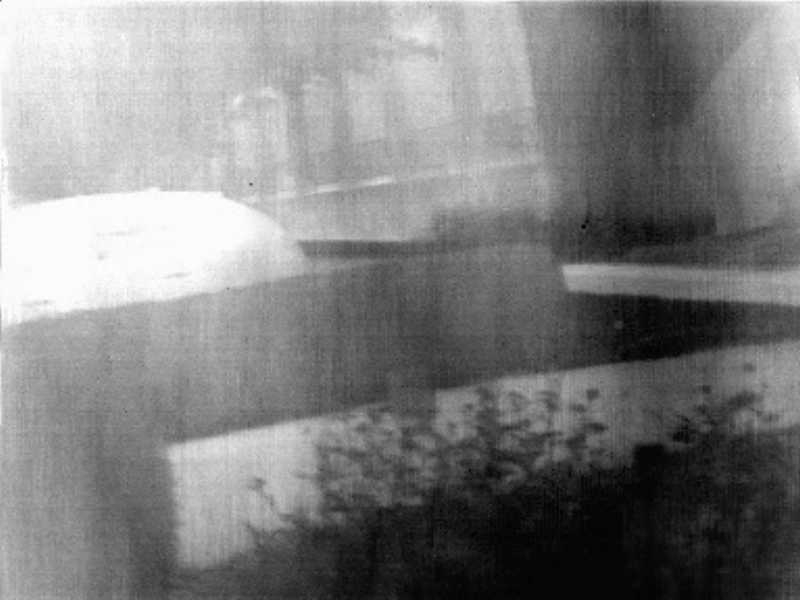}
\caption{\label{fig:expe3} On the left : a real raw image taken by an infrared camera. On the right the result of ADMIRE.}
\end{center}
\end{figure}

\begin{figure}[!h]
\begin{center}
\includegraphics[width=8cm]{./processed/test4}
\includegraphics[width=8cm]{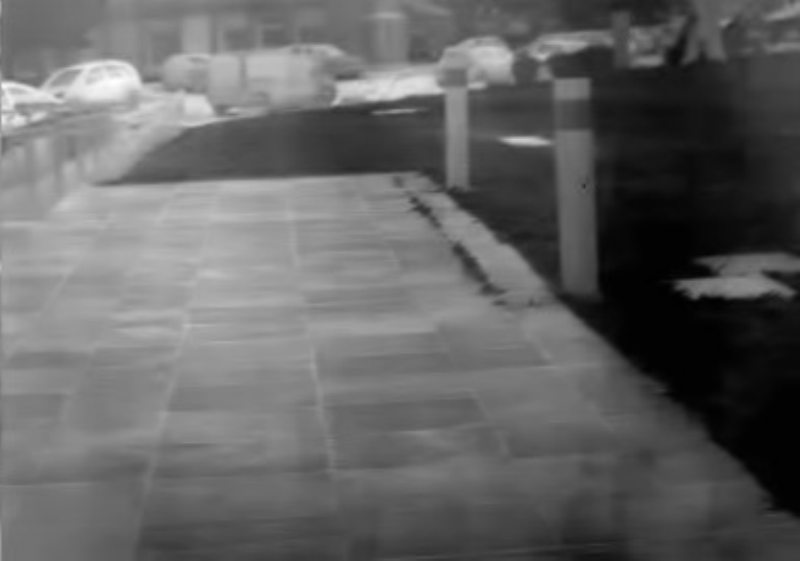}
\caption{\label{fig:expe4} On the left : a real raw image taken by an infrared camera. On the right the result of ADMIRE.}
\end{center}
\end{figure}

\begin{figure}[!h]
\begin{center}
\includegraphics[width=8cm]{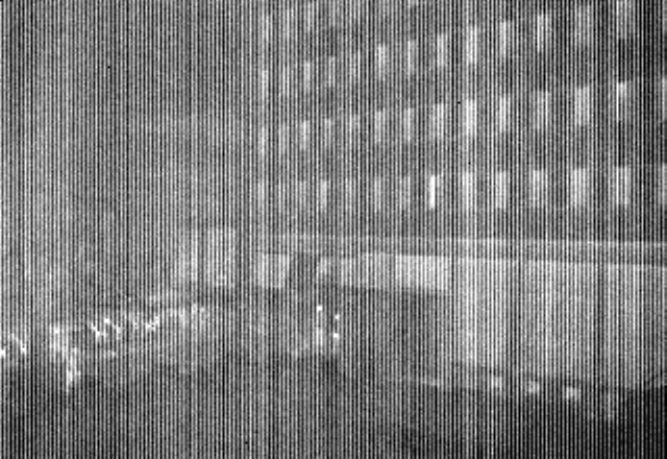}
\includegraphics[width=8cm]{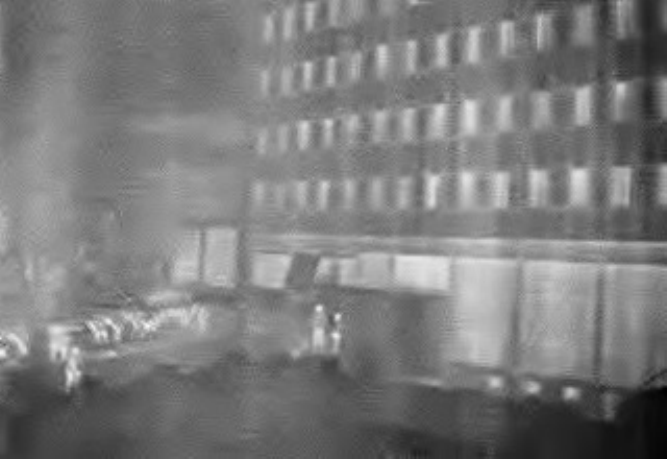}
\caption{\label{fig:expe5} On the left : a real raw image taken by an infrared camera. On the right the result of ADMIRE.}
\end{center}
\end{figure}

\begin{figure}[!h]
\begin{center}
\includegraphics[width=8cm]{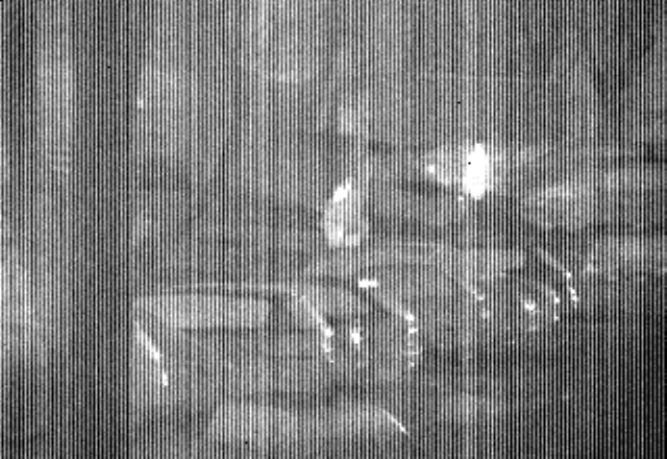}
\includegraphics[width=8cm]{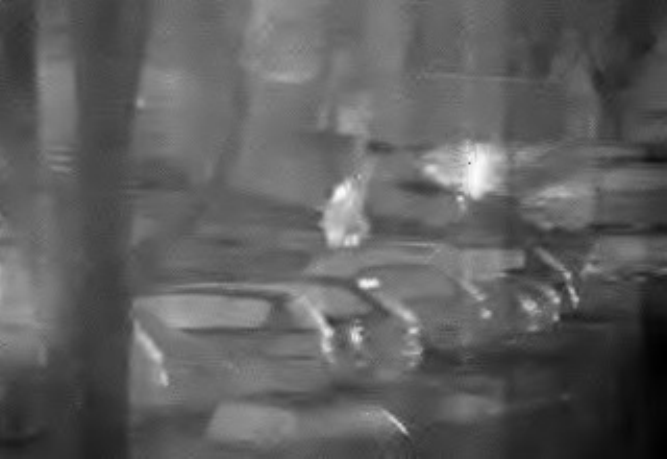}
\caption{\label{fig:expe6} On the left : a real raw image taken by an infrared camera. On the right the result of ADMIRE.}
\end{center}
\end{figure}

The subsequent experiments permits to visualize the result of the proposed method on numerous real raw images. We used different types of landscape to visualize the effect of the proposed non uniformity correction on edges, textures, and  at different level of time exposure (noise). The conclusion is that the quality is quite satisfactory (see Figs.\ref{fig:expe1}-\ref{fig:expe6}).

\section{Discussion and conclusion}
\label{sec:conclusion}
In this paper we start by modeling the image formation chain including the non uniformity and the Poisson (shot) noise terms. From this model we deduce the correct algorithm (chain) to apply to an image in order to perform a good non uniformity correction. We developed  an image processing chain to correct for the non uniformity and the noise. A single image locally adaptive non uniformity correction was designed. It can compensate for fully non linear non uniformity, without any parametric model on the non uniformity side. It does not require motion, or motion compensation, does not need a test pattern or calibration and does not produce any ``ghost artifact''. A state of the art denoising algorithm was modified according to the model to obtain a non uniformity correction chain.  Evaluations using both simulated and real raw images  show that the approach performs an efficient non uniformity correction in term of $RMSE$, contrast invariant $RMSE^{CI}$ and visual image quality. Comparisons were made with a total variation based method. The conclusion is that a single image, ghost-less and non linear non uniformity correction is not only possible but efficient. \\
The next steps are first to obtain a robust noise estimation following, for example, the ``percentile method'' described in \cite{miguel}. Notice that to denoise automatically (without have to tune the $T_i$ and $T_j$ thresholds) theses images a directional estimation of the noise variances are needed. The second step is to take advantage of movies to increase the quality while (still) ensuring the absence of \emph{any} ``ghost artefacts''. Indeed, if needed, the scene change detection is way easier to perform just before the $DCT$ denoising step.

\acknowledgments     

We thank the UCLA math department and the D�l�gation G�n�rale pour l'Armement (DGA) for supporting this work. We are grateful to  St�phane Landeau for providing us the infrared cameras used for the experiments particularly the PEA {\it FUSIBLE}, Image Intensifier/Thermal IR real time fusion prototype, DGA.


\bibliography{biblio2}   
\bibliographystyle{spiebib}   
\end{document}